\documentstyle[graphicx]{mn}

\newif\ifAMStwofonts
\AMStwofontstrue

\def\pg{{PG1211+143}}

\def\einstein{{\it Einstein}}
\def\exosat{{\it EXOSAT}}
\def\xmm{{\it XMM-Newton}}
\def\chandra{{\it Chandra}}

\def\et{{et al.\ }}

\def\ginga{{\it GINGA}}

\def\asca{{\it ASCA}}

\def\xte{{\it RXTE}}

\newcommand{\ls}{\mathrel{\hbox{\rlap{\hbox{\lower4pt\hbox{$\sim$}}}\hbox{$<$}}}}
\newcommand{\gs}{\mathrel{\hbox{\rlap{\hbox{\lower4pt\hbox{$\sim$}}}\hbox{$>$}}}}

\def\arcs{{\hbox{$^{\prime\prime}$}}}

\def\Msun{\hbox{$\rm ~M_{\odot}$}}

\def\H0{{\rm ~km~s^{-1}~Mpc^{-1}}}

\def\msun{M_{\rm \odot}}

\def\et{{et al.}}

\def\deg{^\circ}

\title[A high velocity outflow from \pg]
        {A high velocity ionised outflow and XUV photosphere in the narrow emission line quasar \pg}
\author[K.A.Pounds \et]
        {K.A.Pounds,$^{1}$
	J.N.Reeves,$^{1,2}$
	A.R.King,$^{1}$
	K.L.Page,$^{1}$
	P.T.O'Brien,$^{1}$
         and M.J.L.Turner $^{1}$\\
$^1$ Department of Physics and Astronomy, University of Leicester,
Leicester, LE1 7RH, UK\\
$^2$ Laboratory for High Energy Astrophysics, NASA Goddard Space Flight Center, Greenbelt, MD 20771, USA\\}
\date{Accepted 9 July 2003; Submitted 26 March 2003 ; Revised 8 July 2003}
\pagerange{\pageref{firstpage}--\pageref{lastpage}}
\pubyear{2003}
\begin{document}
\maketitle
\label{firstpage}

\begin{abstract}

We report on the analysis of a $\sim$60 ksec \xmm\ observation of the bright, narrow emission line quasar \pg. 
Absorption lines are seen in both EPIC and RGS spectra corresponding to H- and He-like ions of Fe, S, Mg, Ne, O, N and
C. The observed line energies indicate an ionised outflow velocity of $\sim$24000 km s$^{-1}$. The highest energy
lines require a column density of $N_{H}$$\sim$$5\times10^{23}\rm{cm}^{-2}$, at an ionisation parameter of
log$\xi$$\sim$3.4.  If the origin of this high velocity outflow lies in matter being driven from the inner disc, then
the flow is likely to be optically thick within a radius $\sim$130 Schwarzschild radii, providing a natural explanation for the
Big Blue Bump (and strong soft X-ray) emission in \pg.

\end{abstract}


\section{Introduction}

One of the most striking recent developments in X-ray studies of AGN has been the observation, from high resolution
grating spectra obtained with \chandra\ and \xmm, of complex absorption indicating circumnuclear (often outflowing) matter
existing in a wide range of ionisation states (eg Sako \et\ 2001, Kaspi \et\ 2002). Until recently, however, it has
generally been assumed that this so-called `warm  absorber' was essentially transparent in the `Fe K spectral
band' above $\sim$6 keV, with fluorescent
line emission from the accretion disc being the main feature in AGN spectra at 
those energies (eg Reynolds and Nowak 2002). 

In this paper we report on the spectral analysis of a $\sim$60 ksec \xmm\ observation of the bright quasar \pg. 
At a redshift $z=0.0809$ (Marziani \et\ 1996) \pg\ has a typical X-ray luminosity (2--10~keV) of
$\sim$$10^{44}$~erg s$^{-1}$, for $ H_0 = 75 $~km\,s$^{-1}$\,Mpc$^{-1}$. The Galactic absorption column towards
\pg\ is $N_{H}=2.85\times10^{20}\rm{cm}^{-2}$ (Murphy \et\ 1996), rendering it visible over the whole
($\sim$0.2--12~keV) spectral band of the EPIC and RGS instruments on \xmm.

\pg\ is a low redshift, optically bright quasar, with a strong `Big Blue Bump' (BBB). It is unusual
in the PG sample of bright quasars in having relatively narrow permitted optical emission lines (Boroson and
Green 1992, Kaspi 2000).  \pg\ was first detected in the X-ray band by \einstein, which found a steep
spectrum in the $\sim$0.2-2 keV band (Bechtold \et\ 1997, Elvis \et\ 1991). A subsequent analysis by Saxton \et\
(1993), which combined \exosat\ and \ginga\ data, resolved a strong soft X-ray `excess' above a harder power law
component of photon index $\Gamma$$\sim$2.1. An early \asca\ observation showed the soft excess to be variable,
indicating a source region of $\leq$$10^{15}$cm (Yaqoob \et\ 1994). The improved spectral resolution of \asca\ further 
refined the
broad-band X-ray description of \pg\ (Reeves \et\ 1997), with evidence for a broad Fe K emission line of
equivalent width (EW)$\sim$400-750~eV at $\sim$6.4~keV. A more recent study of the overall (infra-red to X-ray)
spectrum of \pg\ has been published by Janiuk \et\ (2001), considering in particular the strong
emission in the UV and soft X-ray bands and proposing its origin in a warm optically thick `skin' on the
accretion disc. This work also included an analysis of an extended \xte\ observation in 1997, suggesting a cold
reflection factor R~=~$\Omega$/2$\pi$, where $\Omega$ is the solid angle subtended by the reflecting matter, of order unity.

\section{Observation and data reduction}

\pg\ was observed by \xmm\ on 2001 June 15 yielding a useful exposure of $\sim$60 ksec. In this paper we use data
from the EPIC pn camera (Str\"{u}der \et 2001), which has the best sensitivity of any instrument flown to date in
the $\sim$6-10 keV spectral band, the combined EPIC MOS cameras (Turner \et\ 2001), and the Reflection Grating
Spectrometer/RGS (den Herder \et\ 2001). Reference to the Optical Monitor (Mason \et\ 2001) confirmed the strong optical and UV
emission was close to the typical level in \pg. All X-ray data were first screened with the XMM SAS v5.3 software
and events corresponding to patterns 0-4 (single and double pixel events) were selected for the pn data and
patterns 0-12 for MOS1 and MOS2, the latter then being combined. A low energy cut of 200 eV was applied to all
X-ray data and known hot or bad pixels were removed. We extracted source counts within a circular region of
45\arcs\ radius defined around the centroid position of \pg, with the background being taken from a similar
region, offset from but close to the source. The 0.2-10 keV X-ray pn  light curve is reproduced as figure 1 and
shows $\sim$30 percent flux changes over $\sim$6 ksec, similar to those seen in the \asca\ data. Individual
spectra were binned to a minimum of 20 counts per bin, to facilitate use of the $\chi^2$ minimalisation technique
in spectral fitting. Response functions for spectral fitting to the RGS data were generated from the SAS v5.3. 

Spectral fitting was based on the Xspec package (Arnaud 1996) and used a grid of ionised absorber models
calculated with the XSTAR code (Kallman \et\ 1996).  All spectral fits include absorption due to the 
line-of-sight Galactic column of $N_{H}=2.85\times10^{20}\rm{cm}^{-2}$. Errors are quoted at the 90\% confidence
level ($\Delta \chi^{2}=2.7$ for one interesting parameter).

\begin{figure}
\centering
\includegraphics[width=6.3 cm, angle=270]{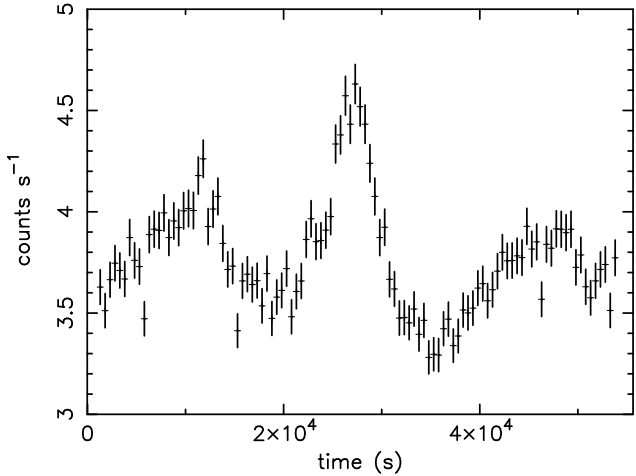}
\caption
{X-ray light curve at 0.2-10 keV from the \xmm\ pn observation of \pg\ on 2001 June 15.}
\end{figure}

\section{1--10 keV spectrum}   

\subsection{Power law} 

X-ray spectra of AGN at 2--10 keV are well fitted, to first order,
with a power law of photon index $\Gamma$ in the range $\sim$1.6-2 for most radio quiet AGN, with a fraction
(eg NLS1) having somewhat steeper indices. The widely held view is that this `hard' X-ray continuum in
Seyfert galaxies arises by Comptonisation of thermal emission from the accretion disc in a `hot' corona (eg
Haardt and Maraschi 1991), and produces additional spectral features by `reflection' from dense matter in the
disc (eg Pounds \et\ 1990, Fabian \et\ 2000).

We began our analysis of \pg\ by confirming there were no obvious spectral changes with source flux
and then
proceeded to fit the \xmm\ pn and MOS data integrated over the full $\sim$60 ksec observation. A simple
power law fit over the 1--10 keV band yielded a photon index of $\Gamma$$\sim$1.79 (pn) and
$\Gamma$$\sim$1.71 (MOS), with a broad excess in the data:model ratio between 3--7 keV, and evidence of
absorption at higher energies in both data sets (figure 2). The fit was statistically unacceptable with an overall
$\chi^{2}$/dof of 1541/1176. When extrapolated to 0.3 keV, the 1--10 keV fits to both pn and MOS data revealed
a strong `soft excess' (figure 3).

\begin{figure}                                                          
\centering                                                              
\includegraphics[width=6.3cm, angle=270]{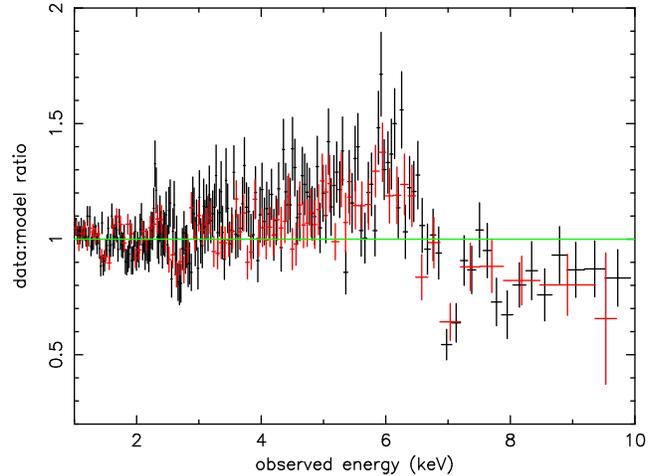}                     
\caption                                                                
{Ratio of the EPIC pn (black) and MOS (red) spectral data to a   
simple power law model fitted between 1-10 keV for \pg.  
The plot shows a broad excess at 3--7 keV and several absorption features including a deep narrow absorption 
line near 7 keV.}      
\end{figure}                                                            
  
\begin{figure}                                                          
\centering                                                              
\includegraphics[width=6.3 cm, angle=270]{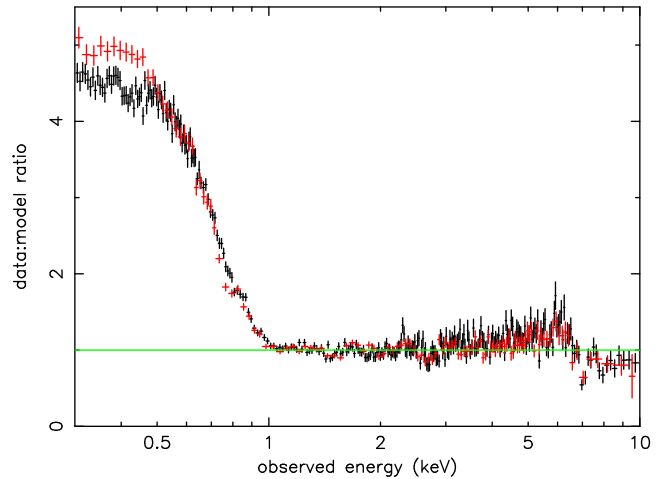}                     
\caption                                                                
{Extension to 0.3 keV of the 1-10 keV power law model fits for the pn (black) and MOS (red) spectral data, 
showing the strong soft excess in \pg.}
\end{figure}      

\subsection{Fe K emission and absorption features}

To improve the 1--10 keV fit we added further spectral components to match the most obvious features in the data. The
indication of an extreme broad emission line suggested reflection from the inner accretion disc, conventionally modelled
with a LAOR line in Xspec (Laor 1991). The addition of a LAOR line, with inclination initially fixed at 30$\deg$ and
$R_{out}$=100$R_{g}$ ( where $R_{\rm g} = GM/c^2$ is the gravitational radius for mass $M$), resulted in a significant
statistical improvement ($\chi^{2}$/dof of 1304/1172), but with an unrealistically large EW of $\sim$1.4 keV (pn) and
$\sim$1.1 keV (MOS). To better fit the broad line profile we added a gaussian line with energy tied to that of the LAOR
line. ( Physically such a gaussian line could represent emission from larger radii on the disc). This  addition gave a
further improvement in the fit, to $\chi^{2}$/dof of 1278/1167. The LAOR line still had a high EW of $\sim$0.6-0.9 keV,
with disc emissivity index $\beta$$\sim$3.5, and inner radius $R_{in}$$\sim$1.5$R_{g}$. The gaussian emission line
component had an rms width $\sigma$ = 0.28$\pm$0.15 keV and EW = 0.25$\pm$0.11 keV.  The (poorly constrained) joint line
energy was  $\sim$6.2 keV, or $\sim$6.7 keV in the source rest frame, implying reflection from ionised matter. 

We then attempted to fit the narrow absorption features visible in figure 2, initially with gaussian shaped absorption  lines
in Xspec. Adding a gaussian line with energy, width and equivalent width free gave a significantly better fit to the absorption
near 7 keV than an absorption edge. The observed line energy was 7.02$\pm$0.03 keV, with $\sigma$$\leq$100 eV, and an
EW of 98$\pm$28 eV. The addition of this gaussian absorption line improved the fit to $\chi^{2}$/dof = 1246/1164. The most
likely identifications of this line are Ly$\alpha$ of FeXXVI or the primary 1s-2p resonance transition in He-like FeXXV. The
rest energies of these lines are separated by 0.26 keV, which would be resolved (or at least produce a broad line) in the EPIC
data. The narrowness of the observed feature at $\sim$7 keV suggests the former identification, with any absorption from the
FeXXV line modifying the Fe K emission line. (We recall evidence for variable line-of-sight absorption superposed on the Fe K
emission line has been previously seen in an \xmm\ observation of Mkn 766; Pounds \et\ 2003b). 

A second narrow gaussian line at 7.9$\pm$0.04 keV was less significant, reducing $\chi^{2}$ to 1234 for 1161 dof. In this case
a statistically better fit ($\chi^{2}$/dof of 1228/1161) was obtained with an absorption edge at $\sim$7.7 keV , or with a
broader line of width $\sigma$ $\sim$0.3 keV centred at $\sim$8.05 keV (EW of 45$\pm$12 eV). We choose to proceed with the latter,
and provisionally identify it with a blend of the FeXXV 1s-3p line and FeXXVI Ly$\beta$, while noting other contributions could
be from absorption edges of less highly ionised Fe (XVII or higher), inner shell transitions as recently addressed by Palmeri
\et\ (2002), or Ni K.  A higher (outflow) velocity component of the absorption line seen at $\sim$7 keV is a further
possibility. 
Figure 2 suggests the presence of other narrow absorption features in the EPIC data, the most significant being
at $\sim$ 2.7 keV, and near 1.5 keV. Fitting these 2 features by successively adding gaussians lines to the model yielded
further reductions in $\chi^{2}$ of, respectively, 26 and 32 for 3 fewer dof in each case (figure 4). 

Details of the absorption lines
thus identified in the EPIC data are summarised in Table 1. When corrected for the redshift of \pg, each line energy indicates
an origin in the same relativistic outflow, with a velocity of $\sim$0.08--0.1c. The best determined line profile,  for
the line at $\sim$7.02 keV, is essentially unresolved, corresponding to a velocity dispersion of $\leq$ 12000 km s$^{-1}$. We
shall see in Section 3.5 that a tighter line width constraint is obtained from the RGS data.

\begin{figure} 
\centering 
\includegraphics[width=6.3 cm, angle=270]{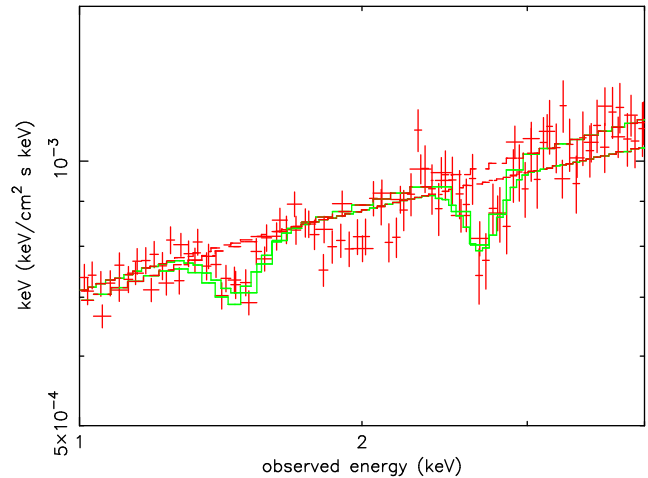} 
\caption {Gaussian absorption line fits to the EPIC pn and MOS data residuals identified with the Ly$\alpha$ lines
of SXVI and MgXII in Table 1. Structure between 1.8--2.2 keV may be related to Si and Au effects in the instrument 
response.}  
\end{figure}

In summary, we find the 1--10 keV spectrum of \pg\ can be described by a broad Fe K emission line,
together with absorption features which are best fitted with gaussian line profiles rather than absorption edges.
The proposed identification of these lines, with resonance absorption from highly ionised Fe, S, and Mg, indicates
an origin in outflowing ionised gas at a velocity of $\sim$0.08--0.1c.

\begin{table*}
\centering
\caption{Absorption lines identified in the parametric fit to the combined EPIC spectrum of \pg. Line energies 
are in keV. The ionisation parameter corresponds to equal abundance of the emitting and recombining ions.}

\begin{tabular}{@{}lccccccc@{}}
\hline
Line  & $E_{obs}$ & $E_{source}$ & $E_{lab}$ &  velocity (km s$^{-1}$) & EW (eV) & log$\xi$  \\

\hline

FeXXVI Ly$\alpha$ & 7.02$\pm$0.03 & 7.59$\pm$0.03 & 6.96 &  24900$\pm$1200 & 98$\pm$28 & 3.8 \\
SXVI Ly$\alpha$ & 2.69$\pm$0.04 & 2.91$\pm$0.04 & 2.62 &  29900$\pm$4000 & 32$\pm$10 & 3.0 \\
MgXII Ly$\alpha$ & 1.49$\pm$0.02 & 1.61$\pm$0.02 & 1.47 &  28600$\pm$4100 & 13$\pm$4 & 2.4 \\

\hline
\end{tabular}
\end{table*}

\subsection{An ionised absorber model}

To quantify the highly ionised matter responsible for the observed absorption features we then replaced the
gaussian absorption lines in the above model with a grid of photoionised absorbers based on the XSTAR code. These model
absorbers cover
a wide range of column density and ionisation parameter, with outflow (or inflow) velocities as a variable parameter.
All abundant elements from C to Fe are included with the relative
abundances as a variable input parameter. In order to limit processing time the fits assume a fixed width of each
absorption line of 1000 km s$^{-1}$ FWHM.  An absorber with ionisation parameter of log$\xi$ = log($L/nr^2$) $\sim$3.4,
and column density of $N_{H}$ $\sim$$5\times10^{23}\rm{cm}^{-2}$, for solar abundances, was found to reproduce the
observed absorption lines at $\sim$7 keV, $\sim$8 keV and $\sim$2.7 keV (figure 5), assuming their indicated identifications,
and an outflow velocity of $\sim$0.08c. We note that most of the
uncertainty  in the derived column density is on the upside, since allowance for partial covering and saturation
in the relatively narrow line profiles would both increase the above value. The feature attributed to MgXII was not
well modelled by the log$\xi$ $\sim$3.4 absorber,
indicating the presence of additional absorbing matter at a lower level of ionisation. This suggests that corresponding
features should be evident in the \xmm\ RGS spectrum. We examine that prediction in Section 3.5.

\begin{figure} 
\centering 
\includegraphics[width=6.3 cm, angle=270]{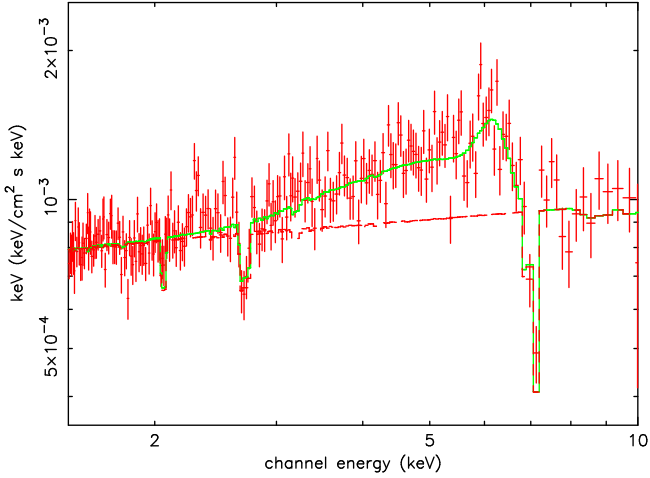} 
\caption {Unfolded spectrum
illustrating the XSTAR fit to the \xmm\ observation of \pg\ as detailed in Section 3.3. For clarity we only
show the pn data.}  
\end{figure}

\subsection{Soft Excess}

Extending the 1--10 keV power law fits for both pn and MOS spectral data to 0.3 keV shows very clearly (figure 3)  the
strong soft excess first indicated by \einstein\ and \exosat\ observations (Elvis \et 1991, Saxton \et 1993). In the
EPIC data this can be adequately modelled with the addition of black body emission components together with 2
surrogate absorption edges. The parameters for \pg\ are a primary blackbody of kT$\sim$110 eV with a weaker component
of $kT$$\sim$265 eV, and `edges' at $\sim$0.8 and $\sim$0.98 keV. We note that a similar combination of
black body emission superimposed by absorption edges has been found previously to be a good parameterisation of EPIC
spectra of AGN (eg. Pounds \et\ 2003a), though - interestingly - in the present case the edge energies are some 10
percent higher than if associated with the OVII and OVIII edges (at rest). Based on this fit we obtain an  average
0.3--10 keV flux for \pg\ of $7.5\times10^{-12}$~erg s$^{-1}$ cm$^{-2}$, corresponding to a luminosity of $\sim
10^{44}$~erg s$^{-1}$ ($ H_0 = 75 $~km\,s$^{-1}$\,Mpc$^{-1}$). The blackbody component is dominant in the 0.3--1 keV
band, representing $\sim$75 percent of the total flux in that band.  The 2--10 keV flux was $2.7\times10^{-12}$~erg
s$^{-1}$ cm$^{-2}$, with a corresponding luminosity of  $3.3\times 10^{43}$~erg s$^{-1}$.

\subsection{Absorption lines in the RGS spectrum}

The strongest absorption lines in the EPIC spectrum have been identified with Lyman alpha of FeXXVI and SXVI. 
The ionisation parameters where these ions and their (recombining) parent ions are in approximate balance are, respectively,
log$\xi$$\sim$3.8 and log$\xi$$\sim$3.0, which essentially determines the effective
high ionisation parameter in the relevant XSTAR fit. The detection of additional absorption in the EPIC data,
including an absorption line attributed to MgXII at $1.47$ keV, suggests the ionised outflow includes matter over
a range of ionisation states, which should be evident in the RGS spectra of \pg. To check this, we began by jointly 
fitting
the RGS-1 and RGS-2 data with a power law and black body continuum (from the EPIC fit) and
examining the residuals by eye. The most obvious spectral features were found to be in absorption, as is usually the
case with Seyfert 1 spectra, a broad line tentatively identified with the forbidden line of OVII (observed at
$\sim$23.8 Angstrom) being the most obvious emission line. A number of weak absorption features are
probably due to Fe L shell absorption, but we concentrate here on the relatively unambiguous identifications
associated with H- and He-like resonance absorption in C, N, O and Ne, since these offer a direct confirmation of
the ionised outflow seen in the EPIC spectrum. Figures 6--9 show the combined RGS1 and RGS2 spectra with the best
fit XSTAR model superimposed. We retained the high ionisation absorber fitted to the EPIC data in the XSTAR model,
and added an intermediate ionisation component to better reproduce the observed absorption in C, N, O and Ne. A
third, low ionisation component was added to match the Fe L edge near 17 Angstrom. The best-fit parameters of
this model were: 

(1) $N_{H}=5\times10^{23}\rm{cm}^{-2}$ at an ionisation parameter of log$\xi$$\sim$3.4; 

(2) $N_{H}=6\times10^{21}\rm{cm}^{-2}$ at an ionisation parameter of log$\xi$$\sim$1.7;

(3) $N_{H}=8\times10^{22}\rm{cm}^{-2}$ at an ionisation parameter of log$\xi$$\sim$-0.9. 
 
Importantly, a large outflow velocity was confirmed from this fit with both of the highly ionised components yielding a velocity of
$\sim$ 24000 km s$^{-1}$. The ionisation parameter at which each detected ion has a similar abundance to its parent ion in a photoionised
plasma is listed in Tables 1 and 2 (Kallman and McCray, 1982), and range from log$\xi$=1.7--2.2, close to the second
outflow component in our XSTAR model fit. Clearly, a single zone photoionised absorber is unable to explain both EPIC and RGS
spectra. The consistent outflow velocities suggest 
a more complex ionisation structure may be due to density variations across the flow.

A visual examination of the RGS spectrum was then carried out to determine the individual line energies, check the line
identifications, and hence the deduced outflow velocities. Although quite deep, the individual line profiles are not very
well determined  due to the limited statistics in the \pg\ data, and this is reflected in the estimated uncertainty in
the line equivalent widths. Nevertheless, we conclude there is no doubt on the identification of the  listed lines and
their consistent `blueshift'. The results of this visual check, summarised in Table 2, yield a weighted mean outflow
velocity of 23400 km s$^{-1}$. As noted above, the individual line profiles are not well determined, but are clearly
narrow. A `combined' Lyman alpha profile from the RGS data, shown in figure 10,
has a measured FWHM of $\sim$2000 km s$^{-1}$, or $\leq$1000 km s$^{-1}$ after allowing for the RGS resolution.
Comparison with the measured outflow velocity suggests the material is streaming outward with relatively little
turbulence, and that we are viewing down (rather than across) the flow. 

The only obvious emission feature in the RGS spectrum, observed at $\sim$23.8 A ($\lambda$$_{lab}$=22.1 A), is most  likely due to
OVII forbidden line emission dispersed across the outflow. Unlike the absorption lines, this line is resolved in the RGS data, with
FWHM $\sim$6000 km s$^{-1}$. The measured EW is 165$\pm$30 mA, a part of which may be due to OVII resonance and intercombination
line emission. The observed flux in the OVII emission line ( $6\times10^{-5}$~photon s$^{-1}$ cm$^{-2}$) is $\sim$2.5
times greater than that in the OVII resonance absorption line. As noted elsewhere the latter may be saturated in the core of the
line, but the strength of the emission line strongly suggests the outflow has a large covering factor, ie the cone angle is wide.
The measured  profile and low projected outflow velocity for the OVII emission line gives further support to such a geometry for
the outflow. Assuming $T_{e}$$\sim 10^{6}$K, and a recombination coefficient (to OVII) of $\sim$$10^{-12}$cm$ ^{3}$ s$^{-1}$  
(Verner and Ferland 1995), the observed OVII emission corresponds to an emission measure ($n^2 V$) of $\sim$$10^{63}$ cm$^{-3}$
for a solar abundance of oxygen and fractional ion abundance of 0.5. Combining this emission measure with the line-of sight column
density ($n\delta^r$)= 5$\times 10^{23}$ cm$^{-2}$ allows an estimate of the radius for a hemispherical outflow (with assumed 
filling factor 0.1)
of $\sim$$3\times$$10^{15}$ cm, with $n$$\sim$$3\times$$10^{8}$cm$^{-3}$. Formally these estimates are upper and lower limits,
respectively, but they are consistent with the product of nr$^{2}$ derived in Section 4.1 from the highest ionisation parameter
in the XSTAR fit. Also, the dominance of the forbidden emission line in the OVII triplet indicates $n$$\leq$ $10^{10}$cm$^{-3}$ 
(Bautista and Kallman 2000).

\begin{table*}
\centering
\caption{Absorption lines identified in the RGS spectrum of \pg. All wavelengths are in Angstroms. The ionisation parameter 
corresponds to equal abundance of the emitting and recombining ions.}

\begin{tabular}{@{}lcccccc@{}}
\hline
Line & $\lambda$$_{obs}$ & $\lambda$$_{source}$ & $\lambda$$_{lab}$ &  velocity km s$^{-1}$ & EW (mA) & log$\xi$ \\

\hline

NeX Ly$\alpha$ & 12.07 $\pm$0.03 & 11.17 & 12.13 & 23700$\pm$800 & 50$\pm$20 & 2.2  \\
NeIX 1s-2p & 13.40 $\pm$0.05 & 12.40 & 13.45 & 23400$\pm$1100 & 70$\pm$15 & 1.8 \\
OVIII Ly$\beta$ & 15.98 $\pm$0.07 & 14.78 & 16.01 & 23000$\pm$1300 & 60$\pm$25 & 1.9 \\
OVIII Ly$\alpha$ & 18.90 $\pm$0.03 & 17.49 & 18.97 & 23400$\pm$470 & 120$\pm$25 &  1.9 \\
OVII 1s-2p & 21.55 $\pm$0.05 & 19.94 & 21.60 & 23100$\pm$700 & 60$\pm$15 & 1.7 \\
OVII 1s-3p & 18.60 $\pm$0.05 & 17.21 & 18.63 & 22900$\pm$810 & 25$\pm$10 & 1.7 \\
NVII Ly$\alpha$ & 24.78$\pm$0.03 & 22.93 & 24.78 & 22400$\pm$360 & 50 $\pm$10 & 1.8 \\
CVI Ly$\alpha$ & 33.62$\pm$0.03 & 31.10 & 33.72 & 23300$\pm$270 & 90$\pm$25 & 1.7 \\

\hline
\end{tabular}
\end{table*}

\begin{figure} 
\centering 
\includegraphics[width=6.3 cm, angle=270]{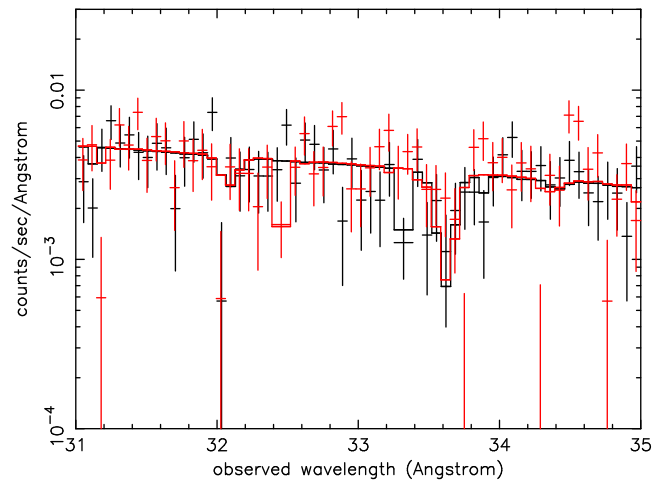} 
\caption 
{RGS spectrum from 31--35
Angstrom fitted with the photoionised model described in Section 3.5. The CVI Ly$\alpha$  absorption line is
observed at 33.62 Angstrom.} 
\end{figure}                                                                                

\begin{figure}
\centering
\includegraphics[width=6.3 cm, angle=270]{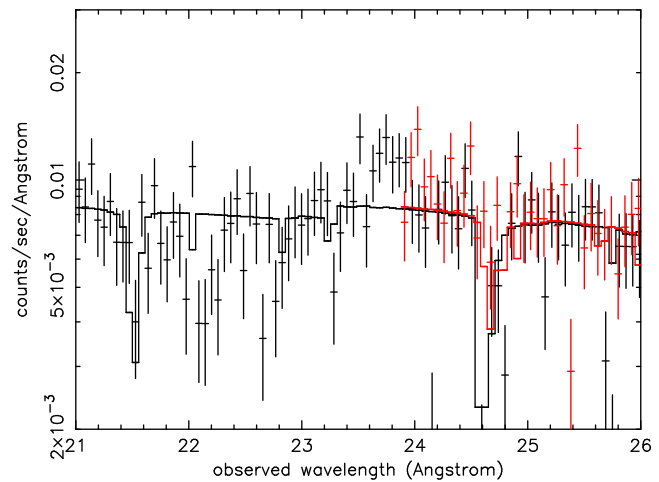}
\caption
{RGS spectrum from 21--26 Angstrom showing resonance absorption lines at 21.55 Angstrom (OVII 1s-2p) and 24.78 
Angstrom (NVII Ly$\alpha$). The relatively strong absorption feature at $\sim$22.1 A is not identified but may be a blend
of the OVII intercombination line pair and the OVI 1s-2p inner shell transition. A broad emission feature at $\sim$23.8 Angstrom is identified with the OVII forbidden
line.} 
\end{figure}   
  
\begin{figure}
\centering
\includegraphics[width=6.3 cm, angle=270]{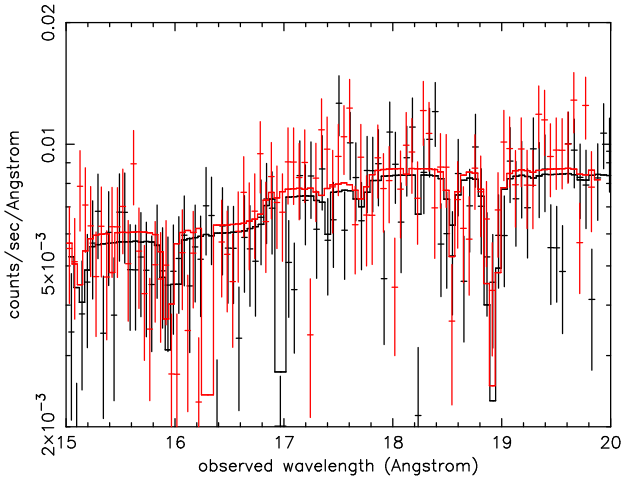}
\caption
{RGS spectrum from 15--20 Angstrom showing resonance absorption lines at 18.90 Angstrom (OVIII Ly$\alpha$), 
18.60 Angstrom (OVII 1s-3p) and 15.98 Angstrom (OVIII Ly$\beta$)} 
\end{figure}     

\begin{figure}
\centering
\includegraphics[width=6.3 cm, angle=270]{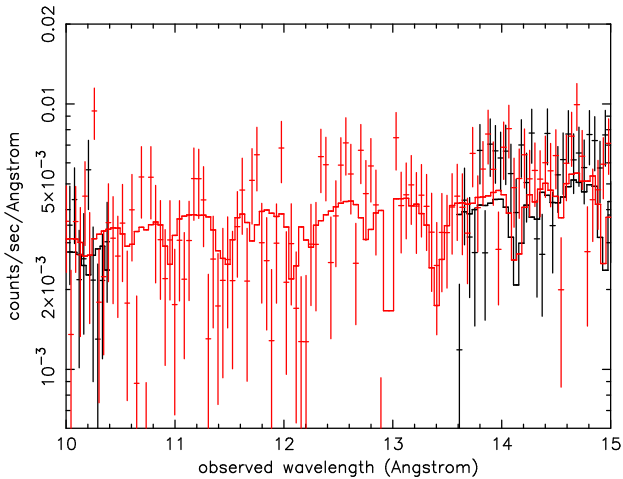}
\caption
{RGS spectrum from 10-15 Angstrom showing absorption lines at 13.40 Angstrom (NeIX 1s-2p) and 12.17 
Angstrom (NeX Ly$\alpha$). Several Fe L lines are also indicated and we note that both Ne lines are probably blended with 
lines of Fe XVII-XXI, limiting their present value in characterising the outflow from \pg} 
\end{figure} 

\begin{figure}
\centering
\includegraphics[width=6.3 cm, angle=270]{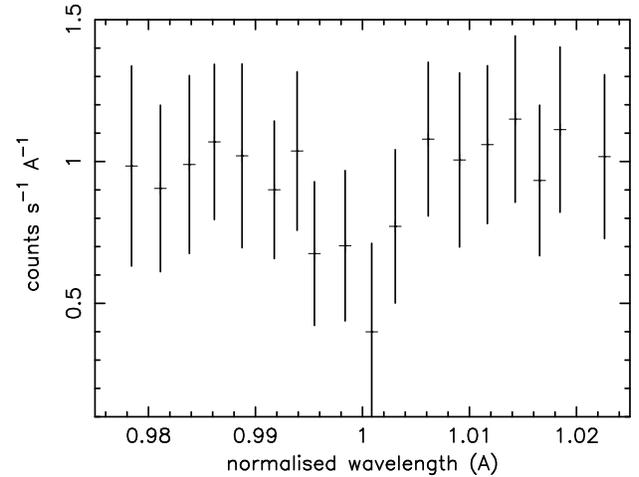}
\caption
{A composite profile of the Ly$\alpha$ lines of CVI, NVII, OVIII and NeX showing the relatively narrow line width
discussed in Section 3.5.} 
\end{figure} 
  
\section{Discussion}

Analysis of the \xmm\ observation of \pg\ has revealed several remarkable features. 

The unusually strong soft excess indicated in previous observations is confirmed. The dominance of this component
below $\sim$ 1 keV suggests an origin as intrinsic thermal emission from the accretion disc, though it has long been
known that the temperature of a standard `thin disc' is too low in AGN to radiate strongly in the X-ray band.
Comptonisation of cooler disc photons in a warm `skin' on the disc surface has previously been invoked as an
explanation of the soft X-ray excess in \pg\ (eg Janiuk \et\ 2001), while Bechtold \et\ (1987) first suggested the
soft X-ray flux was a physical extension of the BBB bump, which is particularly strong in \pg\ and contains much of
the bolometric luminosity. We suggest, in Section 4.2, that this dominant `thermal emission' is a natural consequence 
of the high density outflow.

A second notable feature in the EPIC spectrum is the broad Fe K line emission, exhibiting an extreme `red wing'. We
note that the extreme parameters of this emission line, including the equivalent width, are reduced (but not
removed) when absorption visible in the 7--10 keV band is accounted for. We suggest a possible alternative to the
`relativistic' Fe K emission line in Section 4.3.

The most interesting revelation in the \xmm\ observation of \pg\ is the discovery of an absorption line structure, in both  EPIC and RGS
data, indicating a high column, high ionisation absorber outflowing at a velocity of $\sim$24000 km s$^{-1}$. The remarkable agreement of
the implied velocities from a wide range of absorption lines leaves little doubt that they have a common origin, although the
co-existence of ions with ionisation energies as disparate as FeXXVI and CVI implies a range of ionisation parameter. Although the
coincidence of the measured outflow with the redshift of \pg\ is also remarkable, and we note that \pg\ is viewed through the Magellanic
Stream and the Virgo Cluster (eg Impey \et\ 1999), the presence of a large column density of highly ionised gas in either location would
be a major surprise. The detection of OVII emission at a velocity closer to the systemic velocity of \pg\ further supports the
ionised matter being intrinsic to the AGN. For the remainder of this paper we therefore continue to consider our results in terms of
absorption in \pg. 

\subsection{A high velocity ionised outflow.}

Previous high resolution X-ray spectra of Seyfert 1 galaxies have found a broad range of (low to
moderate) ionisation states and outflow velocities of typically 100--1000 km s$^{-1}$. The long
\chandra\ exposure of the Seyfert galaxy NGC 3783 is a template of such studies (Kaspi \et\ 2002).
However, until now, any absorption in the Fe K band (above $\sim$7 keV) has generally been attributed
to continuum (edge) absorption associated with reflection from the accretion disc. A recent exception
was the report of an absorption feature in the X-ray spectrum of a high redshift BAL quasar (APM
08279+5255), which has been alternatively identified with the absorption edge of Fe XV-XVIII (Hasinger
\et\ 2002), or with strongly blue-shifted resonance absorption lines of Fe XXV or XXVI (Chartas \et\
2002). An earlier \asca\ observation also found evidence for an `absorption line' superimposed on the
`red wing' of a broad Fe K emission line in the Seyfert galaxy NGC 3516 (Nandra \et\ 1999). The
question remains whether highly ionised gas capable of imparting Fe K absorption features on AGN
spectra is a significant component in the outflow of many AGN, and is simply remaining undetected due to
the poor sensitivity of current observations of AGN spectra above $\sim$7 keV. In the latter respect it
may be instructive to note that ionised resonance absorption lines of Fe XXV and XXVI are clearly seen in the
\chandra\ HETGS spectrum of the (much brighter) microquasar GRS 1915+105 (Lee \et\ 2002).

An important aspect of our present observations is that the column density of the most highly ionised matter in the
line-of-sight is high. In the case of \pg\ we find an equivalent hydrogen column (assuming solar abundances) approaching
$10^{24}$~cm$^{-2}$. Although the geometry of the outflow is unknown, it is a reasonable expectation that
near its source the flow is optically thick. The interesting consequence, which we note in Section 4.2, is that the
outflow predicts an inner `photosphere' which is then a natural source of a major part of the radiated luminosity (BBB
to soft X-ray emission) of \pg.

Other important implications following from the detection of a high column, high velocity outflow in \pg\ are a mass loss and kinetic
energy comparable to the accreting mass and the bolometric luminosity. For the observed Fe XXVI absorption line log$\xi$
$\sim$3.8 (Table 1). With an ionising X-ray luminosity ($\ge$7 keV) of 3$\times$$10^{43}$~erg  
s$^{-1}$ we estimate $nr^{2}$
$\sim$$5\times10^{39}$ cm$^{-1}$. Assuming a spherically symmetric flow, at an outflow velocity  of 0.08c, the mass loss rate is then
of order $ \sim 3.5b\msun$~yr$^{-1}$, where $b\leq$1 allows for the collimation of the flow.   Assuming the measured outflow  velocity
is the same as the launch velocity (ie the  material is then `coasting'), the associated kinetic energy is 6.5b$\times$$10^{44}$~erg
s$^{-1}$. 

\subsection{An XUV photosphere}

The previous subsection shows that the column density of the outflow
seen in PG1211+143 is close to being optically thick in the continuum. In
fact this is inevitable if the mass outflow rate is comparable to the
accretion rate required to power radiation at the Eddington limit.

We assume that the outflow quickly reaches a terminal velocity $v$ and
thereafter coasts. Then mass conservation shows that the outflow
density is
\begin{equation}
\rho = {\dot M\over 4\pi vbr^2}
\end{equation}
at radius $r$, where $\dot M$ is the mass loss rate. The electron scattering
optical depth through the outflow, viewed from infinity down to radius
$R$, is
\begin{equation}
\tau = \int_R^\infty\kappa\rho{\rm d}r = {\kappa\dot M\over 4\pi vbR}
\end{equation}
where $\kappa \simeq \sigma_T/m_H$ is the opacity. The Eddington accretion rate is
\begin{equation}
\dot M_{\rm Edd} = {4\pi GM\over \eta \kappa c}
\end{equation}
where $\eta c^2$ is the accretion yield from unit mass. Combining
these equations then gives
\begin{equation}
\tau = {1\over 2\eta b}{R_{\rm s}\over R}{c\over v}{\dot M\over
\dot M_{\rm Edd}}
\end{equation}
where $R_{\rm s} = 2GM/c^2$ is the Schwarzschild radius for mass
$M$. Defining the photospheric radius $R_{\rm ph}$ as the point where $\tau = 1$ gives
\begin{equation}
{R_{\rm ph}\over R_{\rm s}} = {1\over 2\eta b}{c\over v}{\dot M\over \dot
M_{\rm Edd}}\simeq {5\over b}{c\over v}{\dot M\over \dot M_{\rm Edd}}
\end{equation}
where we have taken $\eta \simeq 0.1$ at the last step. Since $b \leq
1, v/c < 1$ we see that $R_{\rm ph} > R_{\rm s}$ for any outflow rate
$\dot M$ of order $\dot M_{\rm Edd}$. In other words, any black hole source accreting above 
the Eddington limit is likely to have a scattering
photosphere at several tens of $R_{\rm s}$. This fact is exploited by Mukai \et\
(2003) and Fabbiano \et\ (2003) in interpreting bright supersoft
sources (including supersoft ULXs), and explored in more detail in a companion paper (King and Pounds 2003).

Since a scattering optical depth $\tau$ degrades photons of energy $\ga500\tau^{-2}$~keV we note the observed
medium-to-hard X-ray flux must be emitted at radii $r \ga R_{\rm ph
}$, and suggest shocks in the outflow as a promising candidate.
This limits the 
medium--energy
X--ray luminosity to a fraction of the outflow kinetic energy $\dot Mv^2/2$, a constraint easily satisfied for \pg. 

In PG1211+143 we estimate $\dot M$/$b$ $ \sim 3.5\msun$~yr$^{-1}$. With $\dot M_{\rm Edd} = 1.6\msun$~yr$^{-1}$,
appropriate for a non-rotating SMBH of $M = 4\times 10^7\msun$ (Kaspi \et\ 2000), we find $R_{\rm ph} \simeq
130R_{\rm s}$ or $1.5\times 10^{15}$~cm.  A wide angle outflow is indicated by the strength and low projected velocity of the
OVII emission line (Section 4.1), corresponding to case 1 in King and Pounds 2003, where the flow geometry limits the 
leakage of photons from the side of the
cone. For \pg\ we therefore assume a value of $b\sim$0.8, which then yields an outflow mass rate of 
$\dot M$ $ \sim 3\msun$~yr$^{-1}$, and a kinetic energy $\sim$ $5\times 10^{44}$~erg~s$^{-1}$. This
is clearly adequate to power a large fraction of the X--ray luminosity $\ga$2 keV of $6\times 10^{43}$~erg~s$^{-1}$

The scattering photosphere discussed above must be larger than the thermalisation region where true absorption
dominates and the soft thermal continuum (BBB) is formed. This implies an effective blackbody temperature 
$\ga 6\times 10^4$~K for a
luminosity of $4\times 10^{45}$~erg~s$^{-1}$, broadly consistent with the strong BBB continuum which dominates the 
bolometric luminosity
of \pg.

\subsection{The relativistic Fe K emission line}
  
An important question raised by the existence of an optically thick photosphere above the inner accretion disc is how the
relativistic Fe K line could be seen, if the innermost accretion disc is obscured.
However, the detection of a large column density of highly ionised matter in the line-of sight to the hard
X-ray source suggests that the extreme `red wing', evident between $\sim$3-6
keV in a simple power law fit to \pg, may actually be an artefact of absorption by more moderately ionised gas partially
covering the X-ray source. Figure 11 shows such an alternative fit to the 1--10 keV EPIC spectrum of \pg. This
fit, with $\Gamma$$\sim$2.04, retains the highly ionised outflow required to model the observed absorption lines, 
but now includes a
second component, of column density  $2\times$$10^{23}$~cm$^{-2}$ and ionisation parameter log$\xi$$\sim$0.5,
with a covering factor of $\sim$0.42. The statistical quality of this fit is comparable to that including the
LAOR line described in Section 3.2. The fit does still require an Fe K emission line, now described by a
Gaussian profile with $\sigma$$\sim$120 eV and EW$\sim$240 eV, at an observed mean energy of $\sim$6.0 keV
($\sim$6.5 keV in the rest frame). We note this emission could arise by reflection from ionised matter in the
unobscured disc, but could also include a significant component via re-emission from ionised gas 
in the outflow, with the forbidden line of FeXXV being a prominent candidate.

\begin{figure}
\centering
\includegraphics[width=6.3 cm, angle=270]{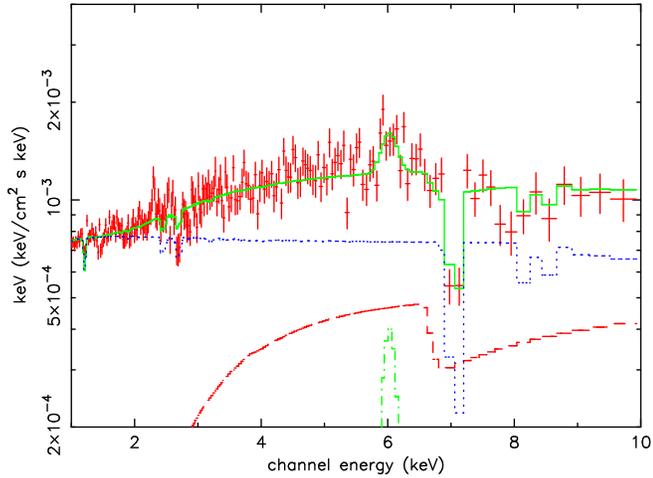}
\caption
{Unfolded spectrum of \pg\ modelled at 1--10 keV with a power law, Gaussian emission line and highly ionised 
outflow. The addition 
of a second, 
lower
ionisation absorber covering $\sim$0.45 of the hard X-ray source replaces the relativistic broad Fe K of earlier 
fits.} 
\end{figure}

\subsection{The X-ray spectrum of \pg, a high accretion rate AGN}

The average 0.3--10~keV band luminosity of \pg\ during our \xmm\ observation was $\sim$$ 10^{44}$~erg s$^{-1}$.
A simultaneous observation with the Optical Monitor on \xmm\ (Mason \et\ 2001) showed the energetically 
dominant BBB to be at a typical value, with a bolometric luminosity of order $4\times$$ 10^{45}$~erg  
s$^{-1}$. Together with the reverberation mass estimate for the SMBH in \pg\ of $M \sim 4 \times 10^{7}\Msun$
(Kaspi \et\ 2000) this luminosity implies accretion in \pg\ at close to the Eddington rate. We have argued 
previously
that a high accretion rate may be the key to the characteristic X-ray properties of Narrow Line Seyfert 1
galaxies (eg Pounds and Vaughan 2000), and the present observation (of a Narrow Line Quasar) suggests a 
further signature of a high accretion rate may appear as Fe K absorption in a highly ionised and massive outflow.

\subsection{Relation to Broad Absorption Line QSOs}

Hitherto most evidence for extreme outflows in AGN has been found in UV sudies of Broad Absorption Line (BAL) QSOs.
The observation reported here, of a massive high velocity outflow from \pg\ broadens the scope of such studies. BAL
QSOs show absorption in a variety of, mainly high-ionisation, UV resonance transitions with velocity widths up to
$\sim 30000$ km s$^{-1}$ (e.g. Weymann \et\ 1991). About 10\% of optically selected QSOs display BALs. As BAL QSOs
appear otherwise similar to non-BAL QSOs, an `orientation model' is traditionally invoked in which BAL QSOs are
those in which the particular line-of-sight intersects an outflow which may be intrinsic to all QSOs. This model has
recently been questioned by the discovery of a relatively high fraction (15 -- 20\%) of radio-loud BAL quasars in
the VLA FIRST survey bright quasar survey (Becker et al. 2000). Becker et al. propose BAL objects may be young or
have recently been fuelled. In any case, the higher fraction of quasars that have BALs implies a higher fraction of
the line-of-sight to the nucleus is covered with substantial absorbing material.

Determining the amount of gas along the line-of-sight to a BAL is generally problematic due to a poor understanding of
the relation between UV and X-ray absorption and the geometry of the flow. Fitting UV absorption lines suggests $N_H \ge
10^{22}$ cm$^{-2}$, whereas the generally weak X-ray fluxes imply columns an order of magnitude or more higher (e.g.
Hamann 1998; Sabra \& Hamann 2001; Gallagher \et\ 2002). Models in which the BAL gas is launched more or less vertically
off a disk and then accelerated by radiation pressure (Murray \et\ 1995; Proga \et\ 2000)  are reasonably consistent
with the UV data but have difficulty in accelerating the large columns of material seen in X-rays unless they are
launched from very close to the black hole  - as we propose for \pg. 

In summary, while PG1211+143 has strong soft X-ray emission and is not a BAL QSO in the UV, it does display a
fast moving outflow and a line-of-sight column density which are similar to those required to explain, respectively, the
UV and X-ray properties of BAL QSOs. Whether the outflow in \pg\ becomes
capable of producing BAL features further out in the flow, but we simply do not intersect such a line of sight, is
unclear. Neither do we yet know how common are X-ray absorption features as reported here for \pg, nor whether the 
X-ray absorbing gas causing BAL QSOs to be `X-ray weak' is in outflow. However, it seems likely
that the BAL phenomena and the high velocity outflow in \pg\ are closely related.

\section{Conclusions}

(1) An \xmm\ observation of the bright quasar \pg\ has revealed evidence of a high velocity ionised outflow,
with a mass and kinetic energy comparable to the accretion mass and bolometric luminosity, respectively.

(2) A further implication of the high observed column density is that the inner flow is likely to be optically thick,
providing a natural explanation for the strong BBB and soft X-ray emission in \pg.

(3) An extreme relativistic Fe K emission line apparent in a simple power law fit to the data can, alternatively,
be explained in terms of partial covering of the continuum source by overlying matter in a lower ionisation
state.

(4) We suggest the above properties might be common in AGN accreting at or close to the Eddington limit.  

\section*{ Acknowledgements }
The results reported here are based on observations obtained with \xmm, an ESA science mission with
instruments and contributions directly funded by ESA Member States and
the USA (NASA).
The authors wish to thank the SOC and SSC teams for organising the \xmm\
observations and initial data reduction and the referee for a careful and constructive reading of the initial text.
ARK gratefully acknowledges a Royal Society Wolfson Research Merit
Award.


\begin{thebibliography}{}
\bibitem{1} Arnaud K.A. \ 1996, ASP Conf. Series, 101, 17
\bibitem{1a} Bautista M.A., Kallman T.R. \ 2000, ApJ, 544, 581
\bibitem{2} Bechtold J., Czerny B., Elvis M., Fabbiano G., Green R.F. \ 1987, ApJ, 314, 699
\bibitem{3} Becker, R.H., White, R.L., Gregg, M.D., Brotherton, M.S., Laurent-Muehleisen, S.A., Arav, N. \ 2000, 
ApJ, 538, 72
\bibitem{3a} Boroson T.A., Green R.F. \ 1992, ApJS, 80, 109
\bibitem{4} Chartas G., Brandt W.N., Gallagher S.C., Garmire G.P. \ 2002, ApJ, 569, 179 
\bibitem{5} den Herder J.W. \et \ 2001,  A\&A, 365, L7
\bibitem{5a} Elvis M., Wilkes B., Giommi P., McDowell J. \ 1991, ApJ, 378, 537
\bibitem{6} Fabbiano, G., Zezas, A., King A.R., Ponman, T.J., Rots, A., Schweizer, F. \ 2003, ApJ, 584, L5
\bibitem{7} Fabian A.C., Iwasawa K., Reynolds C.S., Young A.J. \ 2000, PASP, 112, 1145
\bibitem{9} Gallagher, S.C., Brandt, W.N., Chartas, G., Garmire, G.P. \ 2002, ApJ, 567, 37
\bibitem{10} Haardt F., Maraschi L. \ 1991, ApJ, 350, L81
\bibitem{10a} Hamann, F. \ 1998, ApJ, 500, 798
\bibitem{11} Hasinger G., Schartel N., Komossa S. \ 2002, ApJ, 573, L77
\bibitem{11a} Impey C.D., Petry C.E., Flint K.P. \ 1999, ApJ, 524, 536
\bibitem{11b} Janiuk A., Czerny B., Madejski G.M. \ 2001, ApJ, 557, 408
\bibitem{11c} Kallman T., McCray R. \ 1982, ApJS, 50, 263
\bibitem{11c} Kallman T., Liedahl D., Osterheld A., Goldstein W., Kahn S. \ 1996, ApJ, 465, 994
\bibitem{12} Kaspi S. Smith P.S., Netzer H., Maoz D., Jannuzi B.T., Giveon U. \et. \ 2000, ApJ, 533, 631
\bibitem{13} Kaspi S. \et. \ 2002, ApJ, 574, 643
\bibitem{13a} King A.R., Pounds K.A. \ 2003, MNRAS, in press (astro-ph/0305571)
\bibitem{14} Laor A. \ 1991, ApJ, 376, 90
\bibitem{15} Lee J.C., Reynolds C.S., Remillard R., Shultz N.S., Blackman E.G., Fabian A.C. \ 2002, ApJ, 567, 1102
\bibitem{16} Marziani P., Sulentic J.W., Dultzin-Hacyan D., Clavani M., Moles M. \ 1996, ApJS, 104, 37
\bibitem{16a} Mason K.O. \et \ 2001, A\&A, 365, L36
\bibitem{17} Mukai, K., Pence, W.D., Snowden, S.L., Kuntz, K.D. \ 2003, ApJ, 582, 184
\bibitem{18} Murphy E.M., Lockman F.J., Laor A., Elvis M.\ 1996, ApJS, 105, 369
\bibitem{18a} Murray N., Chiang J.,  Grossman S.A., Voit G.M. \ 1995, ApJ, 451, 498
\bibitem{19a} Nandra K., George I.M., Mushotzky R.F., Turner T.J., Yaqoob T. \ 1999, ApJ, 523, L17 
\bibitem{20} Palmeri P., Mendoza C., Kallman T.R., Bautista M.A. \ 2002, ApJ, 577, L119 
\bibitem{21} Pounds K.A., Nandra K., Stewart G.C., George I.M., Fabian A.C. \ 1990, Nature, 344, 132                                 
\bibitem{22} Pounds K.A., Vaughan S. \ 2000, New Astron. Rev., 44, 431
\bibitem{23} Pounds K.A., Reeves J.N., Page K.L., Edelson R., Matt G., Perola G.C. \ 2003a, MNRAS, 341, 953
\bibitem{23} Pounds K.A., Reeves J.N., Page K.L., Wynn G.A., O'Brien P.T. \ 2003b, MNRAS, 342, 1147
\bibitem{25} Proga D., Stone J.M., Kallman T.R. \ 2000, ApJ, 543, 686
\bibitem{26} Reeves J.N., Turner M.J.L., Ohashi T., Kii T. \ 1997, MNRAS, 292, 468
\bibitem{26b} Reynolds C.S., Nowak M.A. \ 2002, astro-ph/0212065
\bibitem{26c} Sabra, B.M., Hamann, F. \ 2001, ApJ, 563, 555
\bibitem{26e} Sako M. \et\ 2001, A\&A, 365, L168
\bibitem{26f} Saxton R.D., Turner M.J.L., Williams O.R., Stewart G.C., Ohashi T., Kii T. \ 1993, MNRAS, 262, 63
\bibitem{27} Str\"{u}der L.\et \ 2001, A\&A, 365, L18
\bibitem{28} Turner M.J.L. \et \ 2001, A\&A, 365, L27
\bibitem{28a} Verner D.A., Ferland D.J. \ 1996, ApJS, 103, 467
\bibitem{29} Weymann, R.J., Morris, S.L., Foltz, C.B., Hewitt, P.C. \ 1991, ApJ, 373, 23
\bibitem{30} Yaqoob T., Serlemitsos P.,Mushotzky R., Madejski G, Turner T.J., Kunieda H. \ 1994, PASJ, 46, L173
\end{thebibliography}
\end{document}